\newif\ifnotreview
\notreviewtrue

\ifnotreview
	\documentclass[conference]{IEEEtran}
\else
	\documentclass[peerreview]{IEEEtran}
\fi

\usepackage{cite}
\usepackage{amsmath,amssymb,amsfonts}
\usepackage{algorithmic}
\usepackage{graphicx}
\usepackage{textcomp}
\usepackage{xcolor}
\usepackage{flushend}
\usepackage{multicol}
\usepackage{stfloats}
\usepackage{balance}
\usepackage{cuted}
\usepackage{eurosym}
\usepackage{pgffor}
\usepackage{multirow}
\usepackage{subcaption}
\usepackage{booktabs}
\usepackage{tabularx, makecell}

\usepackage[hyphens]{url}
\usepackage[autostyle]{csquotes}
\usepackage[ngerman,british]{babel}

\def\printElphiRetrieved#1#2#3#4#5{
	\begin{subfigure}[c]{\linewidth}
	\foreach \i in {#2, #3, ..., #4}{
		\begin{subfigure}[c]{#5\linewidth}
			\centering
			\includegraphics[width=\textwidth, height=1.5cm,keepaspectratio]{figures/elphi_retrieval/#1_\i.jpg}
		\end{subfigure}
	}
	\end{subfigure}
}

\def\imagesubfigure#1#2{
	\begin{subfigure}{.31\linewidth}
		\centering
		\includegraphics[width=\linewidth]{figures/#1}
		\vspace{-9mm}
		\caption{#2}
	\end{subfigure}
	\hspace{-16mm}
}

\newcommand{\sentimentNegFirst}{Elbphilharmonie \\Expensive, expensive...}
\newcommand{\sentimentNegScnd}{Bad Feng Shui?}
\newcommand{\sentimentNegThird}{Something completely different}

\newcommand{\sentimentPosFirst}{Elbphilharmonie Hafencity Hamburg\\ Even though there were some problems with increasing cost ect. I am sure it will become one of the most fabulous buildings of our time. Amazing architecture, in a wonderful surrounding. So Germany be proud of it.}

\newcommand{\sentimentPosScnd}{Elbphilharmonie - What a beautiful building.}
\newcommand{\sentimentPosThird}{Elbphilharmonie\\ Hamburg harbor birthday from the sailing ship Atlantis}

\newcommand{\sentimentAspectFirst}{Hamburg's Elbphilharmonie concert hall is a spectacular building and architectural masterpiece.}
\newcommand{\sentimentAspectScnd}{Elbphilharmonie (Now one of my favorite buildings)}
\newcommand{\sentimentAspectThird}{Concert and opera house, Elbphilharmonie, in Hamburg.}

\begin{document}
\newcolumntype{P}[1]{>{\raggedright\arraybackslash}p{#1}}

\title{Retrieving Users' Opinions on Social Media with Multimodal Aspect-Based Sentiment Analysis}

\author{\IEEEauthorblockN{Miriam Ansch{\"u}tz}
\IEEEauthorblockA{\textit{Faculty of Informatics} \\
\textit{Technical University of Munich}\\
Munich, Germany \\
miriam.anschuetz@tum.de\\}
\and
\IEEEauthorblockN{Tobias Eder}
\IEEEauthorblockA{\textit{Faculty of Informatics} \\
	\textit{Technical University of Munich}\\
	Munich, Germany \\
	tobias.eder@in.tum.de\\}
\and
\IEEEauthorblockN{Georg Groh}
\IEEEauthorblockA{\textit{Faculty of Informatics} \\
	\textit{Technical University of Munich}\\
	Munich, Germany \\
	grohg@in.tum.de}
}

\ifnotreview
	\maketitle
\else
	\IEEEpeerreviewmaketitle
\fi

\begin{abstract}
People post their opinions and experiences on social media, yielding rich databases of end-users' sentiments. This paper shows to what extent machine learning can analyze and structure these databases. An automated data analysis pipeline is deployed to provide insights into user-generated content for researchers in other domains. First, the domain expert can select an image and a term of interest. Then, the pipeline uses image retrieval to find all images showing similar content and applies aspect-based sentiment analysis to outline users' opinions about the selected term.
As part of an interdisciplinary project between architecture and computer science researchers, an empirical study of Hamburg's Elbphilharmonie was conveyed. Therefore, we selected 300 thousand posts with the hashtag \enquote{\texttt{hamburg}} from the platform Flickr. Image retrieval methods generated a subset of slightly more than 1.5 thousand images displaying the Elbphilharmonie. We found that these posts mainly convey a neutral or positive sentiment towards it. With this pipeline, we suggest a new semantic computing method that offers novel insights into end-users opinions, e.g., for architecture domain experts.
\end{abstract}


\section{Introduction}
Exceptional architecture or star architecture are buildings commissioned for their high recognition value and iconicity. These buildings have a unique design, were designed by a famous architect, or are visually contrasted with their surroundings. They often cause a shift in the scale, spatiality, or content attention about the city they are located in \cite{Sklair-archi-iconicity}.

The propagation of images is a central part of the generation and increase of iconicity. In times of social media, one image can travel around the world within minutes. Therefore, analyzing social media data has become crucial for scientists, e.g., architects who want to investigate the viral effects of a new building \cite{Marti-LBSN-archi-analysis, AlailyMattar-stararchitecture-survey}. On social media platforms, any user can post images of such buildings and express their opinion towards them. This offers the opportunity to retrieve unconstrained opinions by any person, i.e., ones that are not restricted by a questionnaire nor influenced by a biased study design. In addition, multiple user groups can be observed on social media, such as tourists that post about their vacation or local citizens sharing spots in their home city. Another benefit of using social media data for analyzing user opinions is the amount of available data.

However, without the help of automated tools, domain experts have to review and interpret the data manually. When conducting large-scale studies, this results in an infeasible amount of work. Therefore, automated approaches are indispensable for handling large volumes of data. In addition, automated approaches can be re-applied to other datasets or domains, making studies more comparable and reducing the effort even further. A popular automation method is the use of machine learning algorithms. Mainly due to the diversity of textual data, handwritten rules for automation fail to cover the full scope of information in the data.
In contrast, machine learning can capture advanced concepts, e.g., the semantics of texts, or retrieve latent information such as the underlying topic distribution. Therefore, machine learning is proposed for an automated survey on social media data to provide architects access to an amount of data that would otherwise be inaccessible to them. 

\begin{figure}[t]
	\centering
	\includegraphics[width=.36\textwidth]{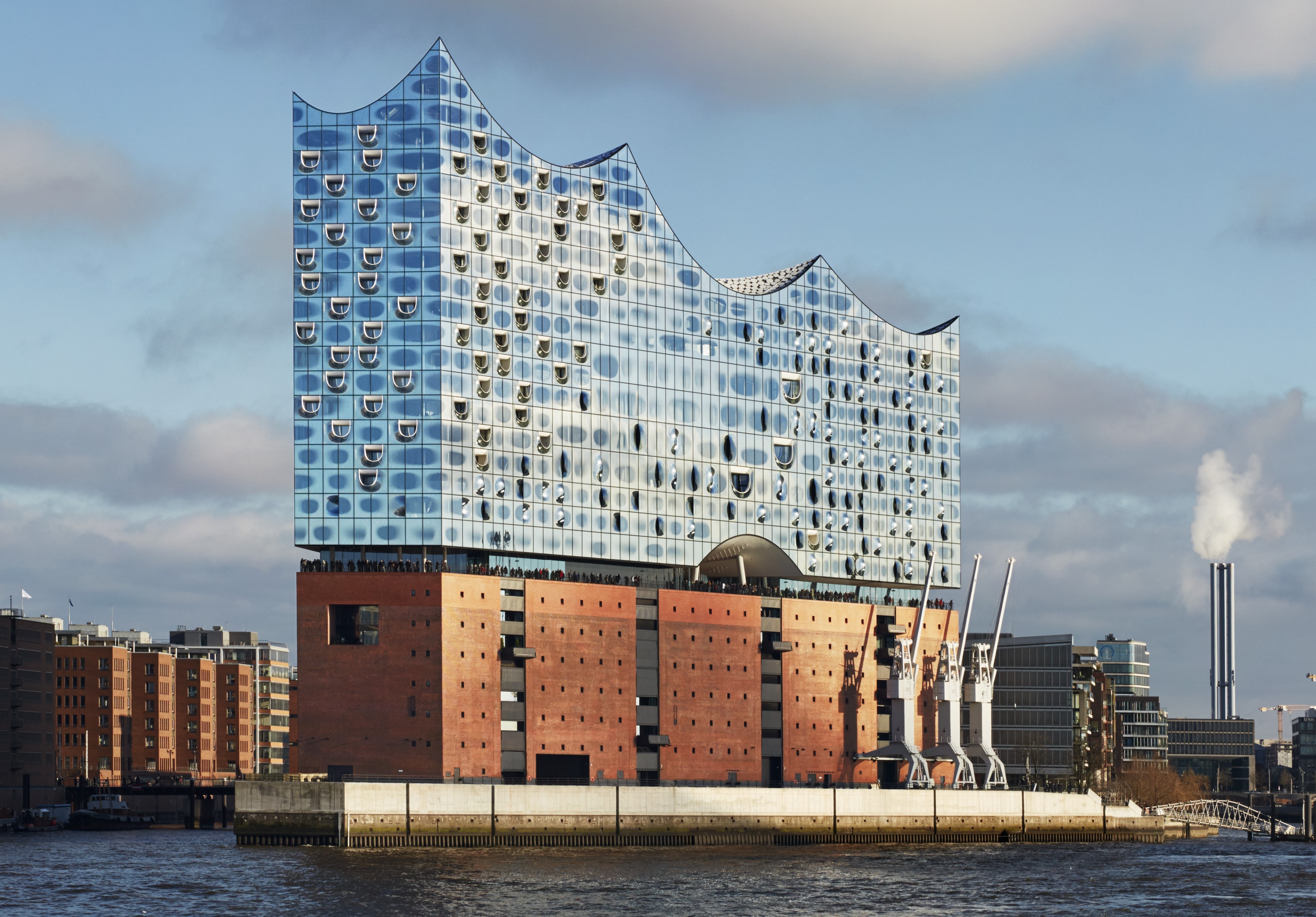}
	\caption[Elbphilharmonie press image $\copyright$ Maxim Schulz]{Elbphilharmonie press image $\copyright$ Maxim Schulz \cite{Elphi-press-image}.}
	\label{fig:elphi_press_foto}
	\vspace{-5mm}
\end{figure}

We attempt to show how machine learning can structure big data and yield interpretations and possible
conclusions based on the data. Therefore, we conveyed an empirical study on social media data to investigate different opinions towards the Elbphilharmonie in Hamburg (see Figure \ref{fig:elphi_press_foto}). The posts about the city of Hamburg were obtained from the image-sharing platform Flickr. The Elbphilharmonie is a philharmonic concert hall by the architecture firm Herzog \& de Meuron, inaugurated in 2017. Hamburg is a Hanseatic city in the north of Germany and has the largest port in its country. Over many years, the city developed northwards, away from the Elbe river and the harbor. In the 1990s, the Hamburg Senate decided to revitalize the former warehouse district at the Elbe riverbank that was abandoned due to the containerization of goods, making the storage capacities of the warehouses superfluous. This revitalization project was called \textit{HafenCity} and yielded a mixed-use urban district with the Elbphilharmonie, the international maritime museum, or the HafenCity university being part of it. The Elbphilharmonie was designed to serve as an icon for this cultural upgrade and a new landmark for Hamburg. The bottom of the building is an old brick warehouse. On top of that is a modern glass construction that imitates a hoisted sail or the sea's waves. This unique architecture stands out. However, many voices were raised that this modern architecture disturbs the view on the historic part of the city. Moreover, the construction cost of the Elbphilharmonie was \euro 866 million, several times more expensive than the initially estimated price. Consequently, the Elbphilharmonie is a controversial building, admired and criticized simultaneously \cite{AlailyMattar-Hamburg}.

This paper aims to provide an overview to the different opinions communicated on social media. The proposed data pipeline processes domain-specific social media data and yields a structured data analysis. 
As part of this pipeline, the domain expert can select an image of a building as the aspect of interest. Then, all images in the dataset depicting the same building are retrieved, and a message-level and aspect-based sentiment analysis is conducted on these posts. Therefore, this contribution is two-fold: On the one hand, a new approach for multimodal aspect-based sentiment analysis on social media data is proposed. On the other hand, this approach was proven effective in an interdisciplinary project between domain experts and computer scientists to conduct an empirical study about the Elbphilharmonie in Hamburg.
The proposed pipeline can be applied to any building by selecting other query images and aspects of interest. In addition, it can be transferred to unseen data by extracting image features from the respective data and updating the queries accordingly. Our code and dataset are published on Github\footnote{\url{https://github.com/MiriUll/multimodal_ABSA_Elbphilharmonie}}.

The remaining paper is structured as follows: Section \ref{sec:rel_work} discusses previous approaches towards image retrieval on landmark images, sentiment analysis on social media, and the combination of both. Section \ref{sec:data} showcases the study design and the resulting dataset from the platform Flickr. Finally, in sections \ref{sec:image_retrieval} and \ref{sec:sentiment}, different image retrieval and sentiment analysis methods are compared on test datasets, and the best-performing ones are applied to filter the Flickr dataset.

\section{Related work}\label{sec:rel_work}
Social media and online review data have been used to mine users' opinions in different domains, for example, opinions towards a specific brand \cite{Mitra-OBIM, Jindal-sentiment-survey}. The methods used in these studies include topic modeling \cite{Liu-B2B-analysis} or aspect-based sentiment analysis \cite{Mitra-OBIM} and focus on textual data. To account for the multimodal nature of social media posts, the authors in \cite{Klostermann-image-text-approach} included the posts' images in their study by clustering them based on their depicted content. As a result, they could identify different brands and products in the images and applied sentiment analysis on the accompanying texts. Similarly, Fang et al. \cite{Fang-multimodal-aspect-opinion} extracted aspects from images, such as buildings, and performed aspect-based sentiment analysis on them. These approaches focus on retrieving different targets addressed in the data. In contrast, the authors in \cite{Rosanensi-Rinjani-mountain} conducted a case study about a specific, pre-defined target, the Rinjani mountain, a popular tourist place in Indonesia. They selected images portraying the mountain in question from social media platforms and performed a dictionary-based sentiment analysis on the image descriptions to retrieve the tourists' opinions towards this destination. This paper reports on a similar study. However, our advanced image-based aspect selection and sentiment analysis approaches yield a more in-depth analysis.
\subsection{Image retrieval on landmark images}
Image retrieval is the task of finding images showing similar objects in an extensive database of images. The images are transformed into a feature space, and the features are compared to find similar contents. In contrast to image classification, where models must classify all images of a class regardless of the intra-class diversity, the image retrieval features must account for precisely these differences \cite{Gordo-deep-image-retrieval}. Traditional techniques, such as scale-invariant features transform (SIFT) \cite{Lowe-SIFT} or KAZE \cite{Alcantarilla-KAZE}, describe images based on distinctive locations and interest points in them \cite{Tareen-comparison-traditional-image-features}. To build a global feature vector based on these local properties, the descriptors are aggregated, e.g., by clustering them into visual words as in the vectors of locally aggregated descriptors (VLAD) \cite{Amato-VLAD}. Other approaches learn image representations with neural networks by fine-tuning pre-trained classification models for the retrieval task. To fine-tune for retrieval on landmark images, the authors in \cite{Noh-DELF-GLD} published the Google landmark dataset and trained their deep local features (DELF) model on it. Other landmark retrieval models are the average precision model \cite{Revaud-Github-model} or the deep local and global features (DELG) model \cite{Cao-DELG}.
\subsection{(Aspect-based) sentiment analysis}
In this paper, two different types of sentiments are analyzed. The message-level sentiment describes the overall sentiment of a post. In contrast, aspect-based models investigate the sentiment about a specific word or phrase in the post. With this, the models can retrieve opinions about a specific topic, independent of the overall sentiment of a message \cite{Cheng-Twitter-Transformer}. Neural networks are a popular method to train classifiers for sentiment prediction because their nested structure can perform an in-depth analysis of the input data and therefore gain a good understanding of complex text features \cite{Dang-message-sentiment-survey}.
The International Workshop on Semantic Evaluation 2017 (SemEval-2017) featured a task about sentiment analysis in Twitter posts \cite{Rosenthal-SemEval2017}. At this task, ensemble models, i.e., models that combine different layer types, were among the most popular and successful competitors. Among them are combinations of convolutional neural networks (CNNs) and long short-term memory networks (LSTMs) \cite{Cliche-BB_twtr, Rouvier-LIA} and a combination of different CNNs \cite{Hamdan-Senti17}. More recently, transformer architectures took over the lead in performance on message-level \cite{Zhang-sent-transformer} and aspect-based classification \cite{Cheng-Twitter-Transformer}. 

\section{Dataset}\label{sec:data}
For this study, social media posts were scraped to form a dataset. Flickr\footnote{\url{https://www.flickr.com/}} is a free online photo-sharing platform similar to Facebook or Instagram.
Although tools like CrowdTangle\footnote{\url{https://www.crowdtangle.com/}} or the official Instagram API\footnote{\url{https://developers.facebook.com/docs/instagram-api/?locale=en_US}} exist to scrape posts from platforms like Facebook and Instagram, these APIs are limited to public posts and large pages with more than 50 thousand followers \cite{CrowdTangle-tracking}.
Since this study was designed to examine the end-user's perspective toward the Elbphilharmonie, the posts by smaller accounts were especially interesting. Moreover, Flickr offers a well-documented API to search and retrieve any public image posted on the platform. Therefore, Flickr was the data source for this study.

Due to the global Covid-pandemic that started in 2020, people's traveling behavior changed dramatically. Therefore, this study only covers the years 2016 to 2019 to avoid having to control for Covid's change of behavior and noise in the data. To retrieve posts related to Hamburg and the Elbphilharmonie, any post with the hashtag \enquote{\texttt{\#hamburg}} was selected. Since not every post with an Elbphilharmonie image mentions it in the text, we selected a broader search tag for better coverage of Elbphilharmonie posts.

The resulting dataset contains 295,387 posts. A post consists of an image and a text body, which again is composed of the image title, a description (both of which can be empty), and a list of hashtags. Many authors upload more than one image simultaneously with a shared text body, i.e., these posts only vary in their image IDs and images but share the same texts and metadata. We call this a \textit{gallery upload}. These gallery uploads can distort spatial and temporal arrangements by over-representing a single user, so they were summarized into one gallery post. The resulting gallery dataset consists of 211,898 posts.

\section{Image retrieval on landmark images}\label{sec:image_retrieval}
The proposed data pipeline used pre-trained models that extracted feature vectors based on the image contents. Then, these vectors were clustered to find images with similar contents. As highlighted in section \ref{sec:rel_work}, multiple local and global feature extraction approaches exist. In this paper, four of them were selected and compared: SIFT \cite{Lowe-SIFT} with VLAD \cite{Amato-VLAD}, DELF \cite{Noh-DELF-GLD}, DELG \cite{Cao-DELG}, and the average precision (AP) model \cite{Revaud-Github-model}.

The scale-invariant feature transform (SIFT) approach detects local maxima in the images and selects some of them as interest points based on a stability measure. Adding gradient information to the keypoints makes them invariant to scale and rotation. These keypoints are local image features, i.e., each encodes a specific image part \cite{Lowe-SIFT}. To combine them into a single, global feature vector, the vector of locally aggregated descriptors (VLAD) \cite{Amato-VLAD} clusters them into visual words. Then, the differences between each descriptor and its closest visual word is calculated, and these differences are concatenated into a global image representation vector.
The deep local features (DELF) are based on a ResNet50 model originally trained for ImageNet classification. These local features are ranked with an attention mechanism, and the top descriptive ones are selected as global image representation \cite{Noh-DELF-GLD}. Both of these approaches, SIFT with VLAD and DELF, extract only local features and concatenate them into a global representation. In contrast, the deep local and global features (DELG) extract both types of features simultaneously. The global features are used for an initial candidate selection of images, while the local features re-rank these candidates based on more specific content analysis \cite{Cao-DELG}. For all three models, the 64 most distinctive features were selected to retrieve vectors of comparable sizes.
Finally, the average precision model generates a fixed-size global feature vector directly. It is based on a ResNet101 architecture and fine-tuned on the Google landmark dataset \cite{Noh-DELF-GLD}. In addition, it was trained using a listwise loss to directly optimize the mean average precision on larger samples of images \cite{Revaud-Github-model, Gordo-deep-image-retrieval}.

\subsection{Image test dataset}
To compare different feature extraction methods, a metric for evaluation is required. The Flickr dataset is unlabeled, i.e., the images are not annotated by the buildings they show. Thus, it is impossible to calculate the retrieval coverage or accuracy directly as the ground truth is missing. Therefore, we selected 36 images from the dataset to build a manually-labeled test dataset. The test images belong to six different classes, for each of which six images were chosen. The six classes include five popular landmarks around Hamburg: the Elbphilharmonie, the Sternschanze, the Holstentor, the Hamburg town hall, and the Hamburg St. Michaelis church. In addition, a random class was added, i.e., random images from the dataset, to include some noise. The selected images represent different levels of difficulties. Each landmark has two very similar images and other images that vary in angle, distance, and cutout of the building. All images are from the Flickr platform and part of the target dataset. The images are unmodified, i.e., they were not cropped nor rescaled. Since the feature extraction is only done once per image and method, the speed of the different methods was no criterion for selection. 
\begin{figure}[b]
	\centering
	\begin{subfigure}[c]{.34\linewidth}
		\includegraphics[width=\textwidth, height=2cm,keepaspectratio]{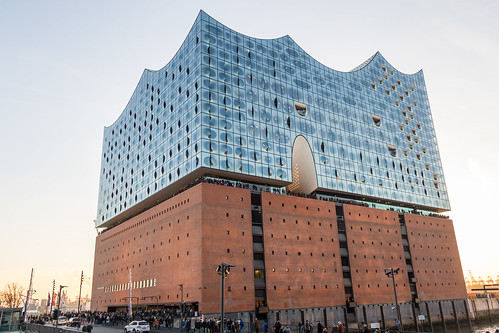}
	\end{subfigure}
	\hspace{-1.5mm}
	\begin{subfigure}[c]{.34\linewidth}
		\includegraphics[width=\textwidth, height=2cm,keepaspectratio]{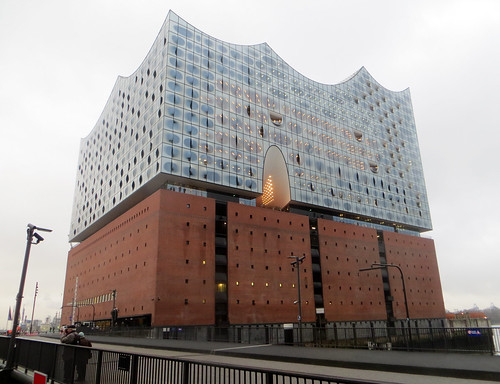}
	\end{subfigure}
	\hspace{-5.5mm}
	\begin{subfigure}[c]{.34\linewidth}
		\includegraphics[width=\textwidth, height=2cm,keepaspectratio]{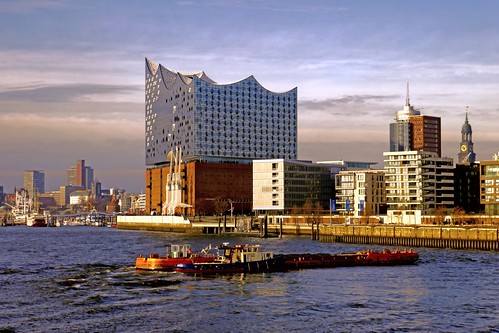}
	\end{subfigure}
	\caption{Image test dataset diversity: Closest and most distant images of the Elbphilharmonie.}
	\label{fig:elphi_img_test}
\end{figure}
\begin{figure*}[ht]
	\hspace{-8mm}
	\imagesubfigure{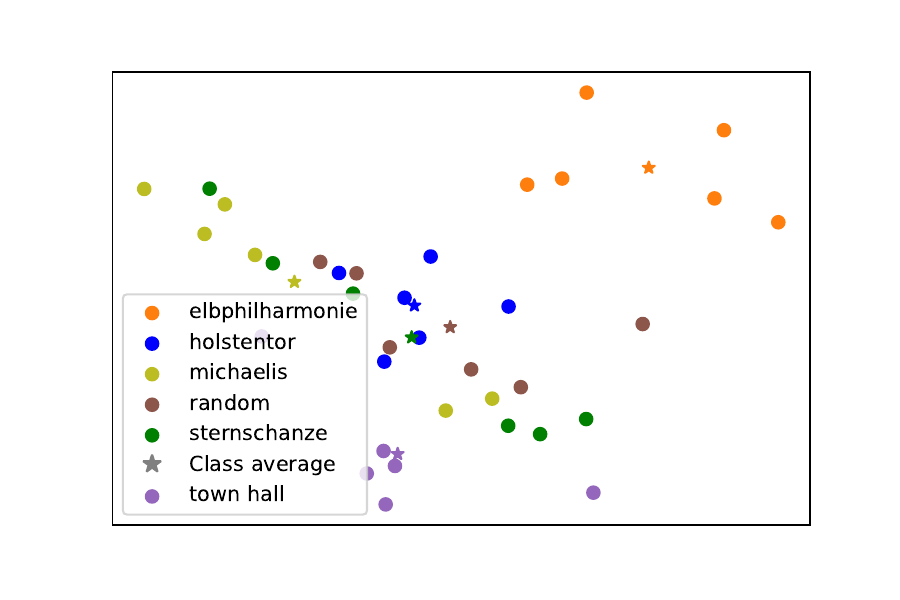}{SIFT with VLAD}
	\imagesubfigure{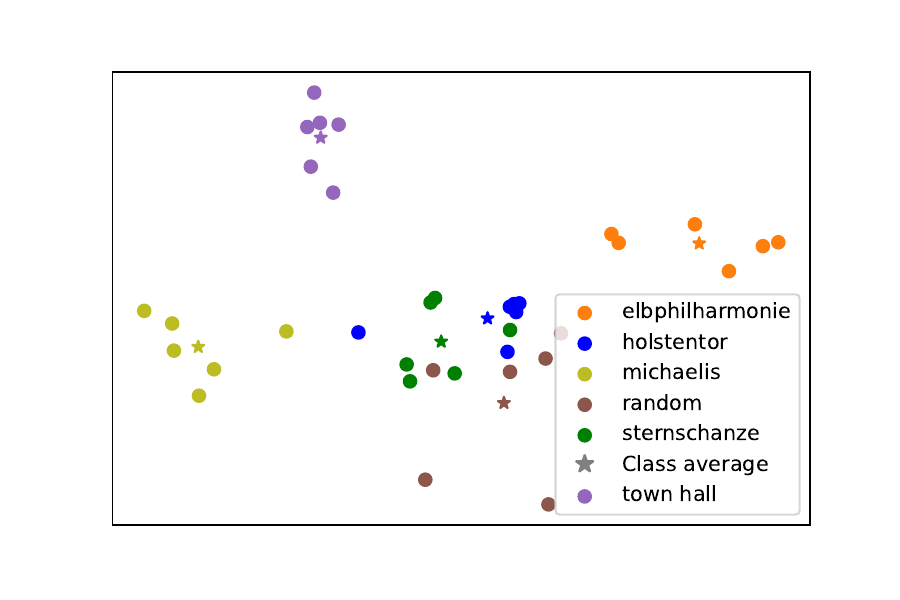}{DELF}
	\imagesubfigure{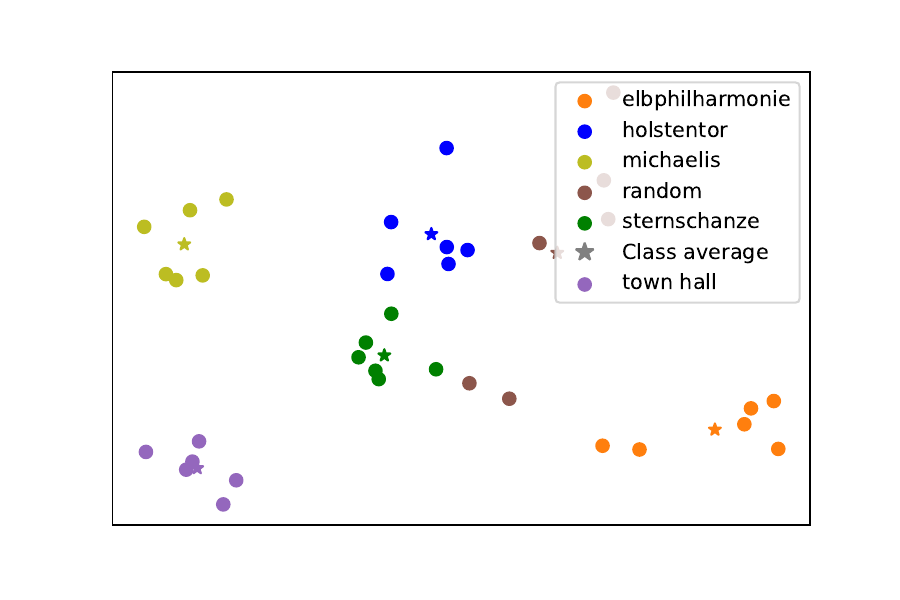}{DELG}
	\imagesubfigure{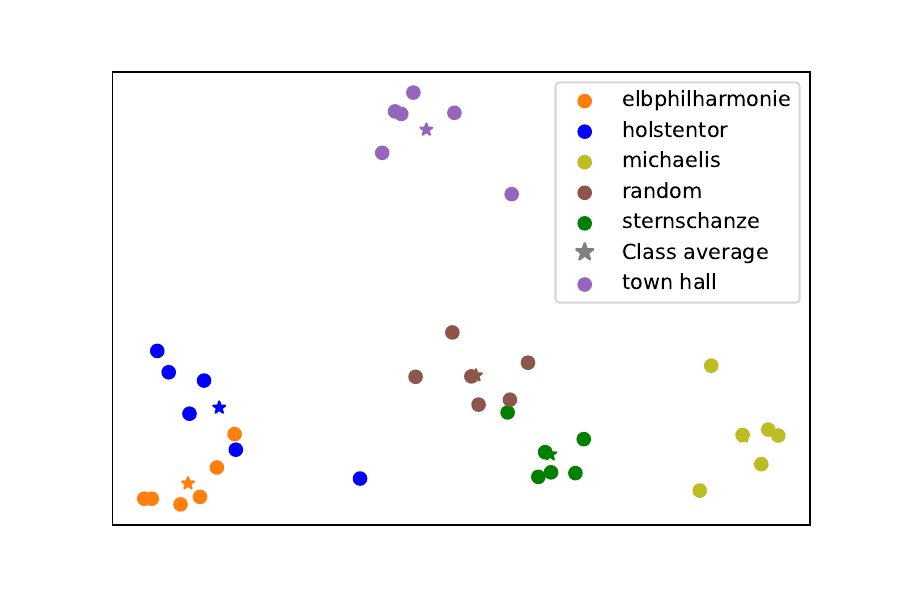}{AP model}
	\caption{Spacial arrangement of PCA projected features based on extracted vectors. For Subfigures (a)-(c), the top 64 descriptors where selected before projecting to two dimensions. Each point represents the feature vector of one specific image. Images that belong to the same classes are colored the same way in all visualizations. A star marker represents the average of the vectors in one class.}
	\label{fig:img_descriptors_pca_vis}
	\vspace{-3mm}
\end{figure*}
In Figure \ref{fig:elphi_img_test}, three test images of the Elbphilharmonie exemplify the diversity in the test dataset. The first two images are very similar and were photographed from the same angle, whereas the third image differs in angle and distance to the building. 

The four selected models extracted feature vectors for each image in the test dataset. Then, we projected these vectors to 2D using principal component analysis (PCA) to visualize their clustering behavior. These projections are shown in Figure \ref{fig:img_descriptors_pca_vis}. All methods can identify clusters for at least some landmark classes, such as the Elbphilharmonie and the town hall. SIFT with VLAD and DELF have a rather dense center with many overlapping classes and, thus, have less discriminative feature vectors.
In contrast, DELG and the AP model show some clear and separated clusters for all classes. For the AP model, the random images seem to build a separate cluster, although they are very different in what they show. A possible reason is that the AP model was explicitly trained for retrieving landmarks. The random images do not show landmarks, so they were clustered into one non-landmark class.

\begin{figure}[h]
	\begin{subfigure}[c]{\linewidth}
		\centering
		\begin{subfigure}[c]{.55\linewidth}
			\centering
			\includegraphics[width=\linewidth, height=4cm,keepaspectratio]{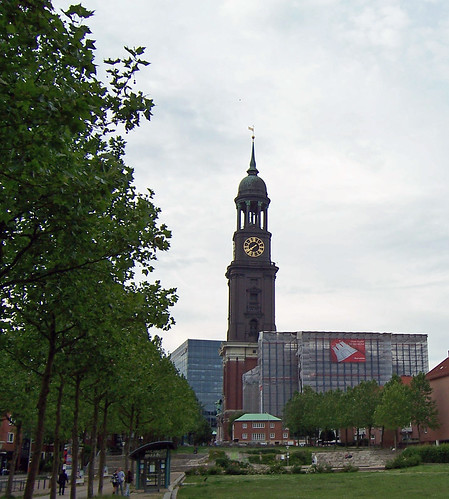}
		\end{subfigure}
		\hspace{-10mm}
		\begin{subfigure}[c]{.39\linewidth}
			\centering
			\includegraphics[width=\linewidth, , height=4cm,keepaspectratio]{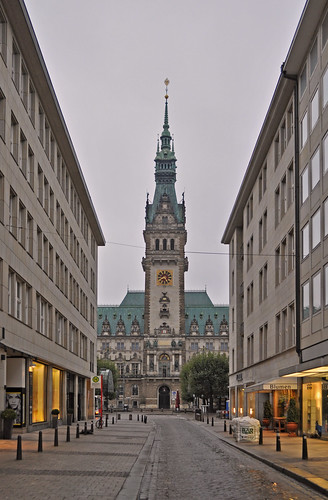}
		\end{subfigure}
		\caption{Michaelis query image and wrongly suggested town hall image}
		\vspace{2mm}
	\end{subfigure}
	\begin{subfigure}[ ]{\linewidth}
		\centering
		\begin{subfigure}{.45\linewidth}
			\includegraphics[width=\linewidth]{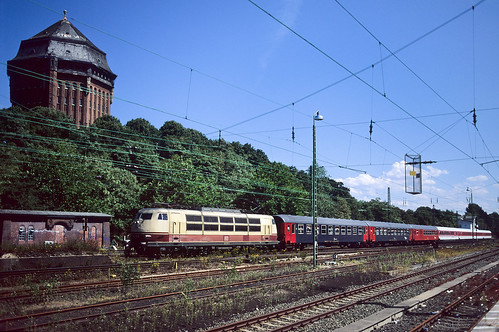}
		\end{subfigure}\vspace{1mm}
		\begin{subfigure}{.45\linewidth}
			\includegraphics[width=\linewidth]{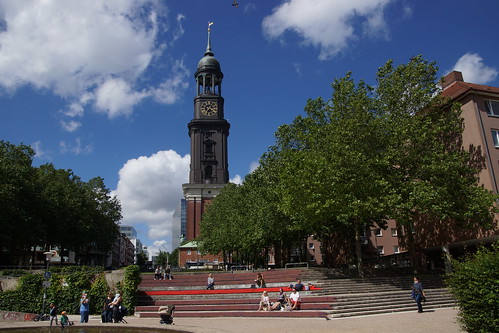}
		\end{subfigure}
		\caption{Sternschanze query image and wrongly suggested Michaelis image}
	\end{subfigure}
	\caption{Images for which the AP model gave wrong suggestions. On the left is the query image and on the right the wrongly suggested image.}
	\label{fig:wrong_ap_images}
\end{figure}

The clusters show that the features help discriminate the different landmark classes. To generate a retrieval model based on these features, a nearest neighbor classifier with an L2-distance was built. This algorithm receives a query image and suggests either the $n$ nearest images or all images within a specified radius in the feature space. All images in the test dataset except the random ones were used as query images. We examined the coverage of correctly suggested images when querying for the three and five closest images that were not the query image itself. Then, the accuracy, i.e., the percentage of images that showed the same landmark as the query image, was calculated. The percentages were averaged over all queries. As the classes are balanced, this average is unbiased. Including the top-3 accuracy accounts for the varying difficulties of the test dataset. Table \ref{tab:test_image_eval} shows the retrieval accuracies on the test dataset. The average precision model achieves a perfect top three accuracy for the feature vectors in the original space. Only for the top five accuracy, it suggested two wrong images. Figure \ref{fig:wrong_ap_images} displays the two pairs of queries and wrongly suggested images. The landmarks in these images were photographed from far away, making it harder to identify the depicted landmark. The remaining four images of the respective class were suggested correctly. The SIFT with VLAD, DELF, and DELG features' performances match the qualities of their clusters, with DELG's top 64 descriptors yielding an outstanding result on the down-projected version of the vectors.
\begin{table}
	\caption{Image test dataset evaluation: Accuracy of retrieved landmark classes. Highlighted are the models with the best score in the respective category.}
	\label{tab:test_image_eval}
	\centering
	\begin{tabular}{lcccc}\toprule
		\multirow{2}{*}{\textbf{Method\textbackslash  Accuracy}} & \multirow{2}{*}{\textbf{Top 3}} & \multirow{2}{*}{\textbf{Top 5}} & \textbf{Top 3} & \textbf{Top 5}\\
		&&&\textbf{PCA} & \textbf{PCA}\\ \midrule
		\textbf{SIFT with VLAD top 64} & 0.62 & 0.53 & 0.53 & 0.48\\
		\textbf{DELF top 64 descriptors} & 0.96 & 0.82 & 0.81 & 0.71 \\
		\textbf{DELG top 64 descriptors} & 0.99 & 0.89 & \textbf{0.94} & 0.\textbf{92}\\
		\textbf{Average precision model} & \textbf{1.0} & \textbf{0.99} & 0.88 & 0.82 \\
		\bottomrule
	\end{tabular}
\end{table}
\subsection{Finding Elbphilharmonie posts}
Based on the performance on the test dataset, the average precision model \cite{Revaud-Github-model} was selected as feature extractor for the full dataset. To retrieve the images with similar content, i.e., to organize the extracted feature vectors, a nearest neighbor model with L2-distance was fit on the data. The query image was the first Elbphilharmonie image from the test dataset (see Figure \ref{fig:elphi_img_test}). The most distant Elbphilharmonie image in the test dataset has an L2 distance of $0.99$ in the feature space. Hence, all images with an L2-distance smaller or equal to $1.0$ were selected from the full dataset. Together with the text bodies of the original posts, they form the Elbphilharmonie-specific dataset. The feature-space distance is a hyperparameter and, thus, needs to be updated when using another query image or searching for different landmarks.

\begin{figure}[ht]
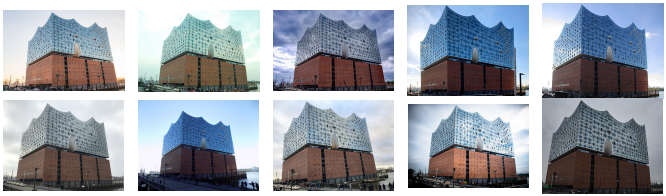
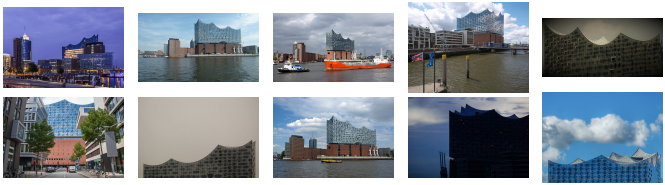

	\begin{subfigure}{\linewidth}
		\printElphiRetrieved{elphi_first}{0}{1}{4}{.181}
		\printElphiRetrieved{elphi_first}{5}{6}{9}{.18}
		\caption{Elbphilharmonie query image and closest images.}
		\vspace{2mm}
		\label{fig:elphi_first_10}
	\end{subfigure}
	\begin{subfigure}{\linewidth}
		\printElphiRetrieved{elphi_last}{0}{1}{4}{.18}
		\vspace{.01cm}
		\printElphiRetrieved{elphi_last}{5}{6}{9}{.18}
		\caption{Most distant Elbphilharmonie images still in radius.}
		\label{fig:elphi_last_10}
	\end{subfigure}
	\caption{Elbphilharmonie images retrieved from full dataset.}
\end{figure}

There were 1,662 images with an L2-distance smaller than one. Figure \ref{fig:elphi_first_10} shows the ten images with the smallest distances. The first image is the query image, i.e., the image with the smallest distance by design. These images show the same view of the Elbphilharmonie and were taken from the same angle and distance to the building. However, they differ in lighting and daytime. In contrast, Figure \ref{fig:elphi_last_10} shows the ten most distant images that still have a distance smaller than $1.0$. These images are very different from the query image, yet, they still show the Elbphilharmonie. Some of them were taken from far away and, thus, display more buildings than only the Elbphilharmonie. Other images show only an excerpt of the building, for example, its wave roof. At first glance, the remaining images seemed to contain a neglectable amount of false positives, i.e., images that do not show the Elbphilharmonie.

To show the benefit of the image-based selection of Elbphilharmonie-related posts, the messages accompanying the images were analyzed towards how many of them mentioned the Elbphilharmonie. The terms \enquote{\texttt{elbphilharmonie}}, \enquote{\texttt{elphi}} or \enquote{\texttt{philharmonie}} were searched in the text bodies, all in upper or lower case. Out of the 1,662 posts showing the Elbphilharmonie, only 1,136 mention it in the text, 953 of them as a hashtag, 263 in the description, and 700 in the image title. Typos, which are relatively common in social media, were not considered in this analysis. Hence, the actual number of mentions may be higher than reported. Nevertheless, 526, i.e., $32\%$, of the posts showing the Elbphilharmonie would have been ignored if the relevant posts had only been selected by text.
\section{Sentiment analysis on Elbphilharmonie posts}\label{sec:sentiment}
The first part of the proposed data analysis pipeline returned a dataset of all images showing the Elbphilharmonie. In the second part, we used this dataset to investigate people's thoughts about this building and in what emotional states the users had taken these images. Therefore, we analyzed the text bodies accompanying the images. As not all texts directly address the Elbphilharmonie, we conducted two different analyses. On the one hand, we examined the overall sentiment of the post, e.g., if someone had a great day in Hamburg and took a photo of the Elbphilharmonie as part of it. Thus, we performed sentiment classification on a message level.
On the other hand, we explored people's comments about the Elbphilharmonie. Hence, aspect-based sentiment analysis inspected the words that occur together with the Elbphilharmonie mentions. Most of the posts in the dataset are written in English or German. Therefore, all of them were translated into English for the sentiment classification part.

Two neural networks were trained, one to predict the post's overall sentiment and one to predict the sentiment of a post towards an aspect within. The SemEval-2017 Task 4 \cite{Rosenthal-SemEval2017} focused on message-level and aspect-based sentiment prediction on Twitter messages. Twitter users post short, informal messages similar to the Flickr text blocks, so the Shared Task's training dataset was considered suitable for our study. The authors published multiple labeled datasets, among them one for classification on a message-level (subtask A) and one for classification of aspect sentiment (subtask C). The first one consists of tweets labeled as positive, negative, or neutral, i.e., it is a three-class dataset. The dataset for subtask C provides tweets together with aspects from the tweets towards which the sentiment was determined. The tweets are assigned to one of these classes: strongly negative, slightly negative, neutral, slightly positive, and strongly positive. As the two classes with a strong sentiment are very small compared to the rest of the dataset, the two sentiment classes with the same polarity were summarized into one class, yielding identical classes as in the first dataset.

To perform the sentiment classifications, the model architectures and preprocessing pipeline of one of the winning teams of the SemEval competition \cite{Baziotis-DataStories} were used. The preprocessing pipeline includes tokenization, segmentation, and spelling corrections. In addition, the authors proposed a bidirectional LSTM network with an attention mechanism for the message-level prediction. Their aspect-based model takes two inputs, the message from which the sentiment is extracted and the aspect words, and feeds them into LSTM networks. These networks have a Siamese structure, i.e., they share their weights in training. Moreover, context-aware attention is used to find the parts of the message that are relevant for the aspect classification. Baziotis et al. \cite{Baziotis-DataStories} reported an accuracy of $0.61$ for the message-level prediction task, which could be reproduced in this work. The aspect-based model achieved a training accuracy of $0.58$ on the simplified, three-class, aspect-based dataset.
\subsection{Sentiment test dataset}
As the Flickr dataset is unlabeled, 13 posts were manually labeled to serve as a test dataset to evaluate the sentiment prediction models. Eight of the posts are targeted toward the Elbphilharmonie, i.e., they explicitly mention it. The spans of words that form the aspects were manually selected. Since no unambiguously negative comment was found in the data, two artificial, negative posts were created. Most of the remaining posts were originally from the Flickr dataset. However, some of them were shortened. Moreover, in some posts, single words were added or deleted. The sentiment test dataset is presented in Table \ref{tab:sentiment_test_comments} in Appendix \ref{app:sentiment_test}.

The message-level model achieves an accuracy of $0.92$ and a macro f1 score of $0.85$ on this test data. It only misclassified one of the negative comments. On the aspect level, only the posts annotated with an aspect-term, towards which the sentiment should be predicted, were selected. The model reaches an accuracy of $1.0$, i.e., all posts were classified correctly. Initially, the aspects in the test dataset were extracted by hand. However, the unlabeled Flickr dataset has no aspect annotations. Thus, the manually selected aspects were compared to the automated aspects \enquote{\texttt{"Elbphilharmonie"}}(with quotation marks included) and \enquote{\texttt{Elbphilharmonie in Hamburg}}. The model's performance remained perfect with these automated aspects.
\subsection{Opinions in Elbphilharmonie posts}
After evaluating the message-level and aspect-based prediction models on a test dataset, they were applied to the Elbphilharmonie dataset posts' text bodies. On a message level, out of the 1,662 posts, four were deemed negative, 113 positive, and the remaining 1,545, i.e., 93\% of the Elbphilharmonie posts, do not carry a sentiment. The aspects in the aspect-based classification must be part of a sentence to predict a sentiment towards it. Therefore, the hashtag lists were omitted for the aspect-based sentiment analysis. Moreover, only the posts in the dataset that mention the Elbphilharmonie in the image title or description were selected. This subset consists of 151 posts. None of these posts was predicted to be negative; however, the model found 16 positive posts. The remaining 135, i.e., 89\% of the posts, do not convey a sentiment towards the Elbphilharmonie. As this study aims to show users' opinions towards the building, the posts with a non-neutral sentiment were inspected. Table \ref{tab:elphi_sentiment_posts} shows examples of posts with positive and negative sentiments as predicted by our models. In the negative posts, users criticize the building costs and address a possible bad feng shui between the Elbphilharmonie and its surroundings. The third negative example describes the Elbphilharmonie as \enquote{something completely different}. It is up to interpretation if this is a negative sentiment or not. In the positive posts, users emphasize the Elbphilharmonie's spectacular architecture and beauty. The first positive message-level example addresses a negative topic, the increased building costs, yet, it has a highly positive sentiment overall. The model could identify this positive sentiment correctly, regardless of the negative aspect. The last example post describes the function of the Elbphilharmonie (a concert hall) rather than conveying a sentiment towards it; thus, it is a neutral comment, but the model failed to identify this.

\begin{table}[t]
	\caption{Example posts from the Elbphilharmonie dataset with different message-level and aspect-based sentiments. The posts' hashtags are omitted.}
	\label{tab:elphi_sentiment_posts}
	\begin{tabularx}{\linewidth}{p{.19\linewidth}p{.11\linewidth}X}
		\toprule
		\textbf{Aspect} & \textbf{Sentiment} & \textbf{Post} \\\midrule
		-- & negative & \makecell[Xt]{\sentimentNegFirst}\\
		-- & negative & \sentimentNegScnd\\
		-- & negative & \makecell[Xt]{\sentimentNegThird}\\
		\midrule
		-- & positive & \makecell[Xt]{\sentimentPosFirst}\\
		-- & positive & \makecell[Xt]{\sentimentPosScnd}\\
		-- & positive & \makecell[Xt]{\sentimentPosThird}\\
		\midrule
		Elbphilharmonie in Hamburg & positive & \sentimentAspectFirst\\
		Elbphilharmonie in Hamburg & positive & \sentimentAspectScnd\\
		Elbphilharmonie in Hamburg & positive & \sentimentAspectThird\\
		\bottomrule
\end{tabularx}
\end{table}

To conclude the sentiment analysis, none of the models were perfect, i.e., false-positive posts were observed in both the message-level and the aspect-based predictions. However, most predictions were correct, and the models could discriminate between the positive and negative classes. By filtering for sentiment-bearing posts, the amount of data was reduced by around 90\%; thus, a more in-depth manual review of the remaining post became feasible.
\section{Discussion}
The data for this study came from the platform Flickr. Most social media users do not use Flickr as their primary platform anymore. Therefore, only opinions from a limited group of people were analyzed and, thus, may induce a bias in the data. Moreover, only a small subset of the dataset was inspected manually. Therefore, the models may have suppressed information or content that could not be considered during interpretation. To counteract these biases, diverse test datasets were generated to evaluate the models' performances for the image retrieval and sentiment analysis pipeline parts. These datasets contain samples of different difficulties, and the models correctly predicted nearly all of them.
Nevertheless, there is no guarantee that the models performed as well on the full dataset. The suggested posts with Elbphilharmonie images were reviewed manually, and only a few false positives were found. The same holds for the sentiment analysis, where most posts predicted to bear a sentiment, in the authors' opinions, did have a sentiment. However, no scan for false negatives was performed on the full dataset. Hence, some Elbphilharmonie-related posts or posts with positive or negative sentiment towards it may be missing.
After all, sentiment is a subjective measure, i.e., people can interpret a statement differently. An example is the description \enquote{something completely different} in one of the example posts. Being different can be deemed both positive or negative depending on the perspective.
\section{Conclusion}
This paper reports the results of an empirical study on how semantic computing can provide insights into user-generated content for domain experts. In addition, we discussed different image-based aspect retrieval and aspect-based sentiment analysis approaches to handle and structure large datasets. Therefore, multiple global image descriptors were compared to find the best image representation for image retrieval on landmark images. The 300 thousand posts in the entire dataset were filtered by relevance, resulting in a dataset with 1,662 images of the Elbphilharmonie.  As indicated by the sentiment analysis models, around 90\% of the users did not express a sentiment towards the Elbphilharmonie. However, those who did were mainly positive, highlighting its exceptional architecture or their fascination with the building. Hence, architects as domain experts can use this preprocessed and analyzed data for further investigation and thus gain access to a substantial amount of data without manual effort. Furthermore, by replacing the query image and aspect, our pipeline can be deployed to any building of interest.
\ifnotreview
\section*{Acknowledgment}
This paper is based on a joined work in the context of Miriam Ansch\"utz's master’s thesis \cite{Anschuetz-MA-stararchitecture}.
\fi

\balance
\flushend
\bibliography{bibliography}
\bibliographystyle{IEEEtran}

\appendices
\section{Sentiment analysis test dataset}\label{app:sentiment_test}
\begin{table*}[b]
	\caption{Posts in the sentiment test dataset. Comments in \textit{italic} were edited (e.g. shortened) or added manually.}
	\label{tab:sentiment_test_comments}
	\centering
	\begin{tabularx}{.9\textwidth}{p{.55\linewidth}p{.1\linewidth}P{.2\linewidth}}
		\toprule
		\textbf{Post} & \textbf{Sentiment} & \textbf{Aspect} \\ \midrule
		A lovely afternoon in Hamburg \#hamburg \#location \#stadt & positive&--\\\hline
		recently I had the chance again to combine job needs and private interests...to stay in Hamburg - i love this city since many years, and I am a ship lover and ship spotter whenever I can: navigare necesse est! \#port \#ship \#harbour \#hamburg \#hafen \#schiff \#elbe \#navigarenecesseest \#elbphilharmonie &positive&--\\\hline
		Hamburg offers a great variety of fascinating old and modern architecture, here are some examples \#hamburg \#elbphilharmonie & positive &-- \\\hline
		On Sunday we had the opportunity to see Hamburg from the water with a friend. It was really great to see everything from the boat. Especially since the port of Flensburg is a bit smaller. \#boat \#sea \#water \#hamburg \#hh \#harbour \#sun \#crane \#ship \#speicherstadt \#elbe \#bridges \#elbphilharmonie \#schumacherphotography \#meineperle & positive &--\\\hline
		Photo tour with Elke in HafenCity in perfect weather :-) & positive&--\\\hline
		\textit{The Elbphilharmonie is a building in Hamburg. It is beautiful.} & positive & The Elbphilharmonie in Hamburg \\\hline
		\textit{View on the beautiful Elbphilharmonie in Hamburg} & positive &Elbphilharmonie in Hamburg\\\hline
		Two Hamburg Icons MS Cap San Diego and the Elbphilharmonie Opera House. While the one is retired and cruises just once a year on the Elbe, the other is not even being finished. & neutral &Elbphilharmonie Opera House\\\hline
		The new opera of Hamburg called "Elbphilharmonie"; built on an old storage building in the seaport. \#opera \#hamburg \#philharmonie & neutral &The new opera of Hamburg called "Elbphilharmonie"\\\hline
		\textit{The building is being built on the former Kaispeicher A, a cocoa, tea and tobacco store. The design and the further structural engineering of the building came from the architects Herzog \& de Meuron. The client is formally the Elbphilharmonie Hamburg Bau GmbH \& Co. KG, whose limited partner and main financier is the City of Hamburg. The Hamburger Prestige object has caused an immense increase in costs over the years. In the basic assessment, costs of 77 million euros were estimated for the Free and Hanseatic City of Hamburg. When the contract was signed in 2007, 114 million euros had already been agreed as the construction cost to be borne by the city.} & neutral &The building The Hamburger Prestige object\\\hline
		\textit{The Elbphilharmonie (also English: Elbe Philharmonic Hall) is a concert hall in the HafenCity quarter of Hamburg, Germany, on the Grasbrook peninsula of the Elbe River. It is one of the largest and most acoustically advanced concert halls in the world. It is popularly nicknamed Elphi.} & neutral &The Elbphilharmonie\\\hline
		\textit{Hamburg's Elbphilharmonie is simply a waste of money} & negative &Hamburg's Elbphilharmonie\\\hline
		\textit{The strange wave roof of the Elbphilharmonie does not fit into the picture of the historic old town. \#defacing \#modern \#ugly} & negative &The strange wave roof of the Elbphilharmonie\\
		\bottomrule
	\end{tabularx}
\end{table*}


\end{document}